\documentclass[aps,twocolumn,superscriptaddress,showpacs,showkeys,amsmath,amssymb,floatfix,nofootinbib]{revtex4}
\usepackage[colorlinks=true,citecolor=blue,filecolor=blue,linkcolor=blue,urlcolor=blue,pdftex]{hyperref}
\usepackage[usenames,dvipsnames]{color}
\usepackage{color}\usepackage{lineno}
\usepackage{graphicx}
\usepackage{bm}
\usepackage[perpage]{footmisc}
\usepackage{amsmath,amsfonts,amssymb,wasysym,ifsym}
\usepackage{color}
\usepackage{MnSymbol}
\usepackage{verbatim}
\usepackage{ulem}
\usepackage{fancyhdr}
\usepackage{datetime}

\newcommand \leff{$\mathcal{L}_{\mathrm{eff}}$}
\newcommand \ly{$L_{\mathrm{y}}$}
\newcommand \qy{$\mathcal{Q}_{\mathrm{y}}$}
\newcommand \kevr{$\mathrm{keV_{nr}}$}
\newcommand \kevee{$\mathrm{keV_{ee}}$}
\newcommand{\xehund}{{XENON100}}

\newcommand{\n}[1]{\mathrm{#1}}
\newcommand{\cSone}{$\mathrm{cS1}$}
\newcommand{\cStwo}{$\mathrm{cS2}$}
\newcommand{\ucSone}{$\mathrm{S1}$}
\newcommand{\ucStwo}{$\mathrm{S2}$}
\newcommand{\Sone}{$\mathrm{S1}$}
\newcommand{\Stwo}{$\mathrm{S2}$}

\begin{document}

\title{Response of the \xehund\ Dark Matter Detector to Nuclear Recoils}

\newcommand{\versionauthors}{2013-02-18}
\newcommand{\versionmembers}{2013-02-08}
\newcommand{\papershorttitle}{Neutron Calibration Paper}


\newcommand{\assergi}{\affiliation{INFN, Laboratori Nazionali del Gran Sasso, Assergi, 67100, Italy}}
\newcommand{\bern}{\affiliation{Albert Einstein Center for Fundamental Physics, University of Bern, Sidlerstrasse 5, 3012 Bern, Switzerland}}
\newcommand{\bologna}{\affiliation{University of Bologna and INFN-Bologna, Bologna, Italy}}
\newcommand{\columbia}{\affiliation{Physics Department, Columbia University, New York, NY 10027, USA}}
\newcommand{\coimbra}{\affiliation{Department of Physics, University of Coimbra, R. Larga, 3004-516, Coimbra, Portugal}}
\newcommand{\heidelberg}{\affiliation{Max-Planck-Institut f\"ur Kernphysik, Saupfercheckweg 1, 69117 Heidelberg, Germany}}
\newcommand{\houston}{\affiliation{Department of Physics and Astronomy, Rice University, Houston, TX 77005 - 1892, USA}}
\newcommand{\laquila}{\affiliation{Department of Physics, University of L'Aquila, 67010, Italy}}
\newcommand{\losangeles}{\affiliation{Physics \& Astronomy Department, University of California, Los Angeles, USA}}
\newcommand{\mainz}{\affiliation{Institute of Physics \& PRISMA Cluster of Excellence, Johannes Gutenberg University Mainz, 55099 Mainz, Germany}}
\newcommand{\munster}{\affiliation{Institut f\"ur Kernphysik, Wilhelms-Universit\"at M\"unster, 48149 M\"unster, Germany}}
\newcommand{\nikhef}{\affiliation{Nikhef  and the University of Amsterdam, Science Park, Amsterdam, Netherlands}}
\newcommand{\purdue}{\affiliation{Department of Physics, Purdue University, West Lafayette, IN 47907, USA}}
\newcommand{\shanghai}{\affiliation{Department of Physics, Shanghai Jiao Tong University, Shanghai, 200240, China}}
\newcommand{\subatech}{\affiliation{SUBATECH, Ecole des Mines de Nantes, CNRS/In2p3, Universit\'e de Nantes, 44307 Nantes, France}}
\newcommand{\torino}{\affiliation{INFN-Torino and Osservatorio Astrofisico di Torino, 10100 Torino, Italy}}
\newcommand{\weizmann}{\affiliation{Department of Particle Physics and Astrophysics, Weizmann Institute of Science, 76100 Rehovot, Israel}}
\newcommand{\zurich}{\affiliation{Physics Institute, University of Z\"{u}rich, Winterthurerstr. 190, CH-8057 Z\"{u}rich, Switzerland}}

\newcommand{\ptb}{\affiliation{Physikalisch-Technische Bundesanstalt (PTB), Bundesallee 100, 38116 Braunschweig,
Germany}\thanks{External collaborating institution}}

\author{E.~Aprile}\columbia 
\author{M.~Alfonsi}\nikhef
\author{K.~Arisaka}\losangeles
\author{F.~Arneodo}\assergi
\author{C.~Balan}\coimbra
\author{L.~Baudis}\zurich
\author{B.~Bauermeister}\mainz
\author{A.~Behrens}\zurich
\author{P.~Beltrame}\weizmann\losangeles
\author{K.~Bokeloh}\munster
\author{A.~Brown}\purdue
\author{E.~Brown}\munster
\author{S.~Bruenner}\heidelberg
\author{G.~Bruno}\assergi
\author{R.~Budnik}\columbia 
\author{J.~M.~R.~Cardoso}\coimbra
\author{W.-T.~Chen}\subatech
\author{B.~Choi}\columbia
\author{A.~P.~Colijn}\nikhef
\author{H.~Contreras}\columbia
\author{J.~P.~Cussonneau}\subatech
\author{M.~P.~Decowski}\nikhef
\author{E.~Duchovni}\weizmann
\author{S.~Fattori}\mainz
\author{A.~D.~Ferella}\assergi\zurich
\author{W.~Fulgione}\torino
\author{F.~Gao}\shanghai
\author{M.~Garbini}\bologna
\author{C.~Geis}\mainz
\author{C.~Ghag}\losangeles
\author{K.-L.~Giboni}\columbia
\author{L.~W.~Goetzke}\columbia
\author{C.~Grignon}\mainz
\author{E.~Gross}\weizmann
\author{W.~Hampel}\heidelberg
\author{R.~Itay}\weizmann
\author{F.~Kaether}\heidelberg
\author{G.~Kessler}\zurich
\author{A.~Kish}\zurich
\author{H.~Landsman}\weizmann
\author{R.~F.~Lang}\purdue
\author{M.~Le~Calloch}\subatech
\author{C.~Levy}\munster
\author{K.~E.~Lim}\columbia
\author{Q.~Lin}\shanghai
\author{S.~Lindemann}\heidelberg
\author{M.~Lindner}\heidelberg
\author{J.~A.~M.~Lopes}\coimbra
\author{K.~Lung}\losangeles
\author{T.~Marrod\'an~Undagoitia}\heidelberg\zurich
\author{F.~V.~Massoli}\bologna
\author{A.~J.~Melgarejo~Fernandez}\columbia
\author{Y.~Meng}\losangeles
\author{M.~Messina}\columbia
\author{A.~Molinario}\torino
\author{K.~Ni}\shanghai
\author{U.~Oberlack}\mainz
\author{S.~E.~A.~Orrigo}\coimbra
\author{E.~Pantic}\losangeles
\author{R.~Persiani}\bologna
\author{G.~Plante}\columbia
\author{N.~Priel}\weizmann
\author{A.~Rizzo}\columbia
\author{S.~Rosendahl}\munster
\author{J.~M.~F.~dos Santos}\coimbra
\author{G.~Sartorelli}\bologna
\author{J.~Schreiner}\heidelberg
\author{M.~Schumann}\bern\zurich
\author{L.~Scotto~Lavina}\subatech
\author{P.~R.~Scovell}\email{paul.scovell@physics.ox.ac.uk}\altaffiliation[present address: ]{Dept of Physics, Univ.\ of Oxford, UK}\losangeles
\author{M.~Selvi}\bologna
\author{P.~Shagin}\houston
\author{H.~Simgen}\heidelberg
\author{A.~Teymourian}\losangeles
\author{D.~Thers}\subatech
\author{O.~Vitells}\weizmann
\author{H.~Wang}\losangeles
\author{M.~Weber}\email{marc.weber@mpi-hd.mpg.de}\heidelberg
\author{C.~Weinheimer}\munster
\collaboration{The XENON100 Collaboration}\noaffiliation
\author{H.~Schuhmacher} \ptb
\author{B.~Wiegel}\ptb


\date{\today}

\begin{abstract}

\noindent Results from the nuclear recoil calibration of the \xehund\ dark matter detector installed underground at the Laboratori Nazionali del Gran Sasso (LNGS), Italy are presented. Data from measurements with an external $^{241}$AmBe neutron source are compared with a detailed Monte Carlo simulation which is used to extract the energy dependent charge-yield \qy\ and relative scintillation efficiency \leff. A very good level of absolute spectral matching is achieved in both observable signal channels -- scintillation \Sone\ and ionization \Stwo\ -- along with agreement in the 2-dimensional particle discrimination space. The results confirm the validity of the derived signal acceptance in earlier reported dark matter searches of the \xehund\ experiment.

\end{abstract}



\maketitle 

\section{INTRODUCTION}

The \xehund~detector~\cite{aprile2011_instrument} aims to detect Galactic dark matter through the elastic scattering of Weakly Interacting Massive Particles (WIMPs) off xenon target nuclei, or (in the absence of signal) to set limits on the WIMP-nucleon interaction cross-section.  
\xehund~is a two-phase (liquid/gas) time projection chamber (TPC) with an active volume containing 62 kg of ultra-pure liquid xenon (LXe), shielded by an active LXe scintillator veto containing 99 kg of the same quality liquid. A total of 242 1 square inch Hamamatsu R8520 photomultiplier tubes (PMTs) are used to read out the two LXe volumes.
The TPC and the active veto are mounted in a low background stainless steel cryostat, enclosed by a passive radiation shield that effectively attenuates and moderates external $\gamma$-ray and neutron background~\cite{aprile2011_instrument}.
The experiment is located underground at the Laboratori Nazionali del Gran Sasso (LNGS), Italy. 
\xehund~simultaneously measures the direct scintillation (\Sone) and the ionization signal, via electroluminescence in the gas (\Stwo), following recoil energy depositions within its active volume.  
This allows event-by-event particle discrimination, and 3-D event vertex reconstruction for definition of a fiducial volume and rejection of events from the outer regions of the TPC.  
Results from 225 live-days exposure have set the most stringent limit on dark matter interactions to date, excluding spin-independent WIMP-nucleon cross-sections above 2~$\times~10^{-45}\,\n{cm}^2$ for a $55~\,\n{GeV}~c^{-2}$ WIMP mass at 90\% confidence level~\cite{aprile2012}.~\xehund\ also excludes spin-dependent WIMP-neutron cross-sections above 3.5~$\times~10^{-40}\,\n{cm}^2$ for a $45~\,\n{GeV}~c^{-2}$ WIMP mass~\cite{aprile2012_SI}.

This article presents results from the comparison between Monte Carlo (MC) simulation of the neutron exposure and the equivalent calibration data.  
Scintillation and ionization channels are assessed independently as well as simultaneously to compare \Sone, \Stwo\ and log$_{10}$(\Stwo/\Sone), the discrimination parameter. 
This assessment allows the parameterization of both the charge yield \qy, the number of ionization electrons produced by a neutron recoil of a given energy, and the relative scintillation efficiency \leff, the energy dependent yield for scintillation photons emitted following a nuclear recoil interaction. 
The two quantities are needed to convert from observed \Sone\ and \Stwo\ signals to true recoil energy deposition, respectively. The ionization and scintillation processes and the correlation between the two are discussed in Ref.~\cite{apriledoke}.
 
Finally, given the derived \leff~and \qy, studies of expected WIMP signals in \xehund\ for two WIMP masses at fixed cross-sections are performed. 

\subsection{Energy Calibration}

The energy scale calibration for electronic recoil interactions in \xehund\ and LXe TPCs in general is determined primarily using known $\gamma$-ray emission lines from standard calibration sources inserted close to the LXe volume.
Energy deposited by the $\gamma$-rays  (through electronic recoils) leads to characteristic features in the physical observable of photoelectrons (PE) in both the \Sone\ and \Stwo\ channels from which the electronic recoil energy scale can be inferred.
This scale is referred to as the electron-equivalent energy scale (\kevee).

In the case of nuclear recoils, the scintillation yield is quenched with respect to that for electron recoils as typically measured at the 122 \kevee\ $\gamma$-ray line of a $^{57}$Co source~\cite{aprile2005}.
This can be parameterized by the relative scintillation efficiency \leff.
The observed \Sone\ must be corrected for any spatially dependent effects (such as variations in light collection efficiency) as detailed in Ref.~\cite{aprile2012_anl}. This corrected value is given the nomenclature \cSone.
The nuclear recoil energy $E$ (in \kevr) is then related to the corrected \Sone\ by:
\begin{equation}\label{leffeq}
E=\frac{\mathrm{cS1}}{L_\mathrm{y}}\frac{1}{\mathcal{L}_{\mathrm{eff}}\left(E\right)} \frac{S_{\mathrm{ee}}}{S_{\mathrm{nr}}}.
\end{equation}
$S_{\mathrm{ee}}$ is the electric field suppression factor for electronic recoils, is measured using the same 122 \kevee\ line and is assumed to be energy independent.
$S_{\mathrm{nr}}$ is the same quantity for nuclear recoils and is also assumed to be energy independent. 
In the \xehund\ detector, operating with a drift field of 0.53 kV~cm$^{-1}$, $S_{\mathrm{nr}}=0.95$ and $S_{\mathrm{ee}}=0.58$~\cite{aprile2006}. 
\ly~is the light yield at operational field for 122~\kevee\ electron recoils and is determined to be (2.28$\pm$0.04) PE/\kevee~for the 225 live-days dark matter search of \xehund~\cite{aprile2012}.

The \Stwo\ signal originates from the ionization electrons produced by particles or radiation interacting in the LXe volume. 
The number of free electrons per unit energy is the charge yield \qy. The electrons drift towards the liquid/gas interface in the presence of a homogeneous electric field of 0.53 kV~cm$^{-1}$. 
The ionization signal is detected as proportional scintillation produced by the extraction and acceleration of electrons in the gas phase above the liquid target volume. A strong electric field of $\sim$12 kV~cm$^{-1}$ gives an extraction efficiency of close to 100\% for an electron reaching the liquid level~\cite{aprile2004,gushchin1979}. Thereby, the conversion from extracted number of electrons to detected S2 photoelectrons is parameterized by the secondary amplification factor, $Y$. As with \Sone, the \Stwo\ signal must be corrected for all spatial effects (such as electron absorption due to the finite electron lifetime in LXe) and is given the nomenclature \cStwo. Summarizing both effects, the nuclear recoil energy $E$ (in \kevr) is related to the ionization signal \cStwo\ (in PE) via the following:

\begin{equation}\label{qyeq}
E=\frac{\mathrm{cS2}}{Y}\frac{1}{\mathcal{Q}_{\mathrm{y}} \left( E \right)}.
\end{equation}
For \xehund, $Y$ has been measured by a dedicated analysis of the single electron \Stwo\ gain. The amplified signal follows a gaussian distribution with a well defined mean of (19.5 $\pm$ 0.1)~PE/e$^{-}$ and 1\,$\sigma$ width of $\sigma_Y = 6.7$~PE/e$^{-}$, valid for the time period of the presented neutron calibration. A detailed study of the single electron response will be presented in an upcoming paper. The conversion between S2 and \kevr\ may be field-dependent (seen in Ref.~\cite{aprile2006}) but sufficient measurements have not been made to fully parameterize this dependence.

The sensitivity of any dark matter detector depends critically on its response to low energy elastic nuclear recoil interactions.  
Typically, broad spectrum neutron sources, such as a $^{241}$AmBe ($\alpha$,n) neutron source, are used for calibrating the detector to such nuclear recoils, as in \xehund.
A method for determining the energy-dependent response of the detector to nuclear recoils is to compare neutron calibration data for a $^{241}$AmBe exposure with a detailed MC simulation. The $^{241}$AmBe calibration data were taken at the beginning of the reported 225 live-days dark matter run.
This method complements direct measurements using dedicated experimental setups with lower mass but higher light and charge yields per unit recoil energy as presented in $e.g.$~Refs.~\cite{plante2011,manzur2010}. 

Historically, the \Sone\ and \Stwo\ channels have been presented independently~\cite{sorensen2009,horn2011}. However, exploiting both \Sone\ and \Stwo\ channels together in data-MC comparison allows considerably more robust constraints. In such a way, the detector response can be mapped both in \cSone, according to the extracted \leff, and \cStwo, with \qy, simultaneously.
Such necessary consistency probes the ability of the MC simulations to reproduce energy dependent event distributions where \Sone\ and \Stwo\ channels are combined to provide discrimination.
Additional verification of the detector response can be achieved through comparison of the simulated source neutron emission rate required for spectral matching. The neutron emission rate used in the simulation must match the independently measured source strength for true agreement between data and MC. 

\section{Modeling Neutron Interactions}
\label{sec:modellingneutrons}

The \xehund\ instrument including the shielding and surrounding environment has been modeled in detail using the GEANT4.9.3 toolkit~\cite{geant4} as previously described in Ref.~\cite{em_background}.  
The physics inputs for this model have been extended to simulate the $^{241}$AmBe neutron calibration exposure of \xehund, with nuclear recoil angular cross-sections calculated using the ENDF/B-VI/B-VII databases~\cite{endf2006} provided in the data files G4NDL~3.13. 
The input $^{241}$AmBe spectrum adopted is that of an ISO 8529-1 standard~\cite{iso}.  
Results are confirmed to be robust against uncertainties in the initial spectral shape since the final recoil spectrum in the active volume depends only weakly on it. Average variations of less than 5\% in the simulated recoil energy spectrum are found and considered sub-dominant to other sources of uncertainty discussed in this publication.  

The neutron emission rate of the $^{241}$AmBe source was measured at the Physikalisch-Technische Bundesanstalt (PTB), the German National Metrology Institute, in August 2012. To measure with high accuracy, the main component of the PTB Bonner sphere neutron spectrometer was used.
The setup is well suited for (n,$\gamma$) discrimination. 
The neutron emission rate of the \xehund\ $^{241}$AmBe source was determined using the ratio of count rates of this source and a reference $^{241}$AmBe source, the rate of which is well known and traceable to national standards. The measurement includes a systematic check of the flux isotropy and results in an integral source strength of $(160 \pm 4)$~n/s.

Coincidences between nuclear recoils in the active LXe volume with direct $\gamma$-rays from the source (such as the 3.2~MeV or 4.4~MeV $\gamma$-rays from de-excitation of $^{12}$C* populated by the Be($\alpha$,n) reaction, as well as a number of low energy $\gamma$-rays that do not reach the sensitive volume of the detector) are negligible as there is no angular correlation between the emitted neutron and the coincident $\gamma$-ray.
In addition, $\gamma$-rays from the $^{241}$AmBe are attenuated through the use of a 10~cm thick lead wall mounted between the source delivery tube and the LXe volume.

With the quoted source strength, pile-up effects from further source related or delayed emission, such as 2.2~MeV $\gamma$-rays following radiative capture of thermal neutrons on the hydrogenous shielding, are also insignificant (according to the simulation).
Excitation lines due to the activation of the xenon by the neutron source give signals at high values of log$_{10}$(\cStwo/\cSone). These are effectively removed by cutting on this discrimination parameter.
In addition, the contribution from scattering to metastable states has been studied (as this may lead to changes in the spectral state) but has been found to be negligible.

Energy depositions from nuclear recoil interactions in the simulation are selected and the process of signal generation, including detector threshold and resolution effects, is modeled as described in the following sub-sections.  
The final selection of single-scatter recoil events includes any multiple scatters that are indistinguishable from single scatters due to proximity of vertices in all three dimensions, or due to small secondary energy depositions that would be sub-threshold in \xehund~\cite{aprile2011_instrument}, and are present at the level of about 1\%.

Since \leff\ has been measured with greater accuracy and down to lower recoil energies in comparison to \qy~\cite{plante2011,manzur2010,horn2011}, the following approach to model the simulated signal response is adopted: As described below, \qy\ is deduced by performing a $\chi^2$-fit of the MC generated \cStwo\ spectrum to data. 
The impact of systematic uncertainties in correlated parameters, such as signal efficiency and \leff, are quantified. For the determination of \qy, a global fit to all previous direct measurements of \leff (as described in Ref~\cite{aprile2011}) is used. Once \qy\ has been obtained, a parameterization of \leff\ is extracted using the same $\chi^2$ minimization technique giving an absolute matching between the \cSone\ data and corresponding MC spectrum. A robust description of both the charge and scintillation yield is achieved and, therefore, it is shown that the measured \cSone\ and \cStwo\ distributions can be matched absolutely to the MC framework using the same inputs and methods as developed for recent dark matter analyses~\cite{aprile2011,aprile2012_anl}.

\subsection{Monte Carlo S2 and S1 Signal Generation}

The signal conversion and detector response is computed based on the energy deposition of nuclear recoil events recorded by the GEANT4 model.

As a first step in generating the eventual \cStwo\ signals, the number of electrons $n_e$ liberated at the event vertex position is simulated. The mean value $\overline{n}_e = \mathcal{Q}_{\mathrm{y}}\,(E)E$ is calculated from the definition of the charge yield function \qy. Poisson fluctuations of this quantity are taken into account on an event-by-event basis. In the subsequent step it is considered that the finite electron lifetime in LXe~\cite{aprile1991} determines the number of electrons that reach the liquid/gas interface after drifting along a uniform electric field. Experimentally, the electron loss can be quantified using the depth-dependent charge-yield distribution of \Stwo\ signals.
For the $^{241}$AmBe calibration run, the electron lifetime was measured using the 40 \kevee\ signal associated with the de-excitation of $^{129}$Xe nuclei following inelastic interactions with neutrons. The exponential suppression factor was found to be $\tau_e = 356\,\mu$s which is consistent with a measurement using a $^{137}$Cs source (as used to determine the electron lifetime throughout the dark matter search run). Both methods combined have a systematic uncertainty of $\sim$2\%.
The simulation takes into account the signal loss as a function of electron drift time $t_d$, hence, $n_e(t_d) = n_e \exp({-t_d/\tau_e})$.
Finally, the measured Gaussian \Stwo\ response of single electrons is applied, described by the mean amplification $Y$ and width $\sigma_Y$, both introduced in the context of Eq.~(\ref{qyeq}), to the number $n_e(t_d)$ of simulated electrons reaching the gas phase. The resulting signal represents the uncorrected \Stwo\ and, hence, has all spatial corrections removed in order to provide a comparison with the raw \Stwo\ (\ucStwo) recorded in \xehund.
The procedure is repeated in the simulation for every nuclear recoil energy deposit recorded along a given neutron track.
Depending on the S2 size and $z$ position of any scatter, the expected S2 width is calculated using a parameterization obtained from actual data. Two neighboring scatters can be resolved if their drift time distance is bigger than twice the average width of both S2 signals. Below this threshold, there is still some finite number of resolvable scatters but the roll-off of that distribution is rather sharp. When averaging over the entire volume this peak separation corresponds to 3mm distance on average. Any unresolvable scatters are summed and constitute around 10\% of the total number of recoils. For comparison with the method used in the analysis of experimental data, the resulting array of uncorrected \ucStwo\ signals is ordered by size, and for each entry the corrected \cStwo\ value is additionally computed by inverting the electron lifetime suppression, thus $\mathrm{cS2} = \mathrm{S2}\,\exp({t_d/\tau_e})$.
\\ \\
For the generation of \Sone, simulated recoil energy is converted to an observable photoelectron signal using Eq. (\ref{leffeq}). The resolution on this average quantity is sampled by assuming a Poisson distribution, which yields an integer number.
However, since the light collection efficiency (LCE) throughout the \xehund\ liquid volume is not uniform, simulated numbers of primary photons are adjusted using the position dependent, experimentally known LCE correction map~\cite{aprile2011_instrument} in order to reproduce the true position-dependent light-yield and derive the detected, i.e. uncorrected, \ucSone\ signal size. Subsequently, the response of the PMT electron amplification to the emission of a single photoelectron from the photocathode is taken into account. The mean signal resolution of the PMTs, determined using an LED calibration to monitor the single-photoelectron response and measured to be 0.5~PE for the 1~PE peak, is applied to add Gaussian smearing to the number of photoelectrons. Finally, the corrected \cSone\ is inferred by applying the experimental LCE map in inverse direction, this time inserting the reconstructed event position of the largest \ucStwo\ signal determined before. In this way the simulation method correctly resembles the processing of actual data.

\subsection{Defining Cuts and Efficiencies in Simulated Data}
\label{cuts}

In order to reliably extract \qy\ and \leff, fiducial, energy and data quality cuts are applied to both data and MC signals. 
A fiducial volume of 34~kg (equivalent to that used in Ref.~\cite{aprile2012}) is selected using both radial and drift-time cuts. 

It is important to measure the response of \xehund\ to single-scatter nuclear recoil events as potential WIMPs interacting in the active volume of the detector will do so only once.
In addition, the \Sone\ light from multiple scatter interactions will, in fact, result in a single pulse. In the multiple-scatter case, the relation between \leff\ and recoil energy in Eq.~(\ref{leffeq}) is inappropriate.
Consequently, single scatter events are selected in both data and MC using identical parameterization as described 
in Ref.~\cite{aprile2012_anl}.

Furthermore, a two-fold PMT coincidence requirement for \Sone\ signals is defined in the data. Two signals larger than 0.3~PE must be seen in separate photomultiplier tubes within a time window of $\pm$20~ns. As photon tracking is not included in the simulation, the energy-dependent efficiency of this cut -- as determined using calibration data~\cite{aprile2012,aprile2012_anl} -- is considered in the MC spectrum. In addition, to avoid threshold effects for small \Stwo\ signals, an \Stwo\ threshold (defined using the uncorrected electron lifetime \ucStwo) of 150~PE is applied to both data and MC. 

\begin{figure}[h]
\includegraphics[width=.5\textwidth]{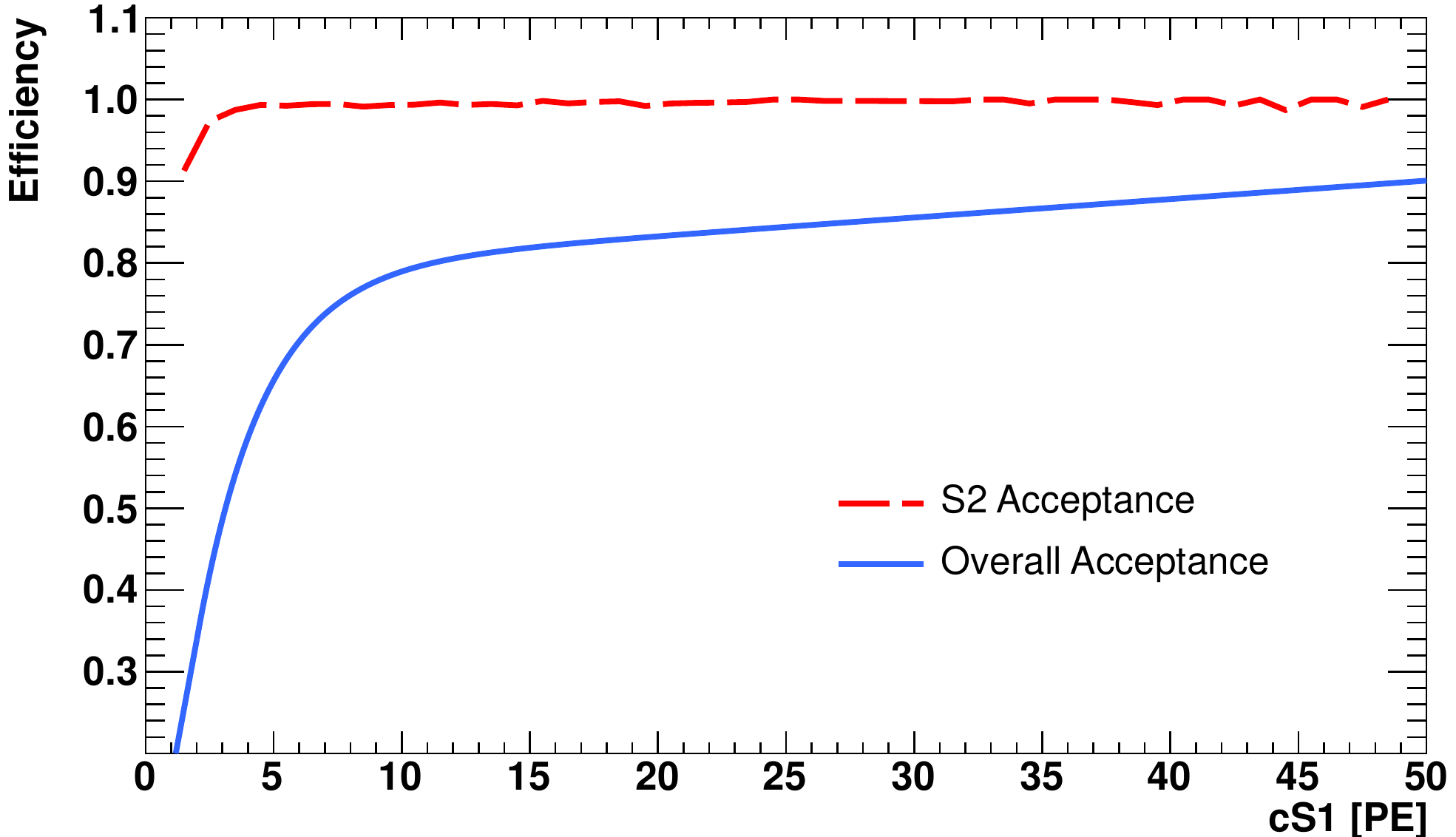}
\caption{\label{acc_pl}  
Cut efficiencies used in this analysis. The red (dashed) line represents the effect of the \Stwo\ threshold efficiency on \cSone. This efficiency is extracted directly from the MC simulation, taking the best fit Qy as input. The blue line represents the efficiency curve when all data quality cuts have been applied. Details of the cut efficiencies used can be found in Ref.~\cite{aprile2012_anl}. In the process of spectral matching, the function is allowed to vary within 10\%.}
\end{figure}

Finally, cuts must be applied to the calibration data to remove spurious events that are accepted as single scatters. No additional noise signals are added to the MC simulation, hence, the efficiency of these cuts as derived using calibration data is applied to the MC spectrum. The definition and energy dependent efficiencies of these cuts are discussed in depth in Ref.~\cite{aprile2012_anl}.

Fig.~\ref{acc_pl} shows the efficiency for the \Stwo\ threshold cut which is extracted directly from the simulation and translated to an efficiency as a function of \cSone. Also shown is the overall efficiency function used in this publication which includes all other cuts mentioned above. 

\section{Method and Results}
\subsection{Ionization Channel -- Determining \qy}
As a first step \qy\ is derived by fitting the simulated \cStwo\ spectrum to the one observed in data. In this process, \leff\ remains fixed to the parameterization presented in Ref.~\cite{aprile2011}. 

A $\chi^2$-minimization technique~\cite{lourakis04LM} is used to find the best matching between data and MC by varying pivot points of an Akima spline~\cite{akima} interpolation of \qy. For every intermediate $\chi^2$ computation, the non-linear descent algorithm requires the re-evaluation of the detector response, applying the updated \qy\ to generate \Stwo. 

\qy\ is parameterized by 8 unconstrained and independent spline pivot-points at 0.5, 3, 8, 15, 25, 40, 100 and 250~\kevr. The lowest pivot point is added to provide an unbiased extrapolation to zero recoil energy but has effectively no impact on the spectral matching. In data, the corrected \cStwo\ spectrum ranges from 0 to 8000~PE, divided into 65 bins of equal width.

The impact of various simulation parameters on the best-fit \qy\ was studied to estimate the systematic uncertainty of the final result.
The largest systematic error is connected to the choice of \leff\ as variations in this quantity lead to changes in the simulated \cSone\ spectrum and, consequently, in the number of events passing the selection requirements. With a lower (higher) value of \leff\ the \cSone\ energy spectrum of accepted events will be shifted upwards (downwards). Accordingly, \qy\ will decrease (increase) in order to compensate this effect and re-establish the matching in \cStwo. This interdependency is present mainly near the detection threshold, where the acceptance as function of \cSone\ falls steeply (Fig.~\ref{acc_pl}), and becomes negligible at higher recoil energies. The \leff\ parameterization is allowed to vary within the $\pm1\sigma$ uncertainty bounds as defined in Ref.~\cite{aprile2011}. Similarly, the \cSone\ efficiency function was allowed to vary by $\pm$10\% around its reported mean. This 10\% variation is a conservative estimate of the uncertainty on the acceptance as the statistical errors on the data-driven cut acceptance are $\sim$2\% with point to point fluctuations of the same size~\cite{aprile2012_anl}. The systematic error connected to the choice of pivot positions and initial values has been found to be negligible in the energy region above 3~\kevr\ (the lowest energy at which \leff\ has been directly measured~\cite{plante2011}). Finally, the statistical uncertainty of about 1\% on average is also included. This is obtained after repeating the simulations about 50 times at fixed configurations but varying random seeds.

\begin{figure}[h]
\includegraphics[width=.5\textwidth]{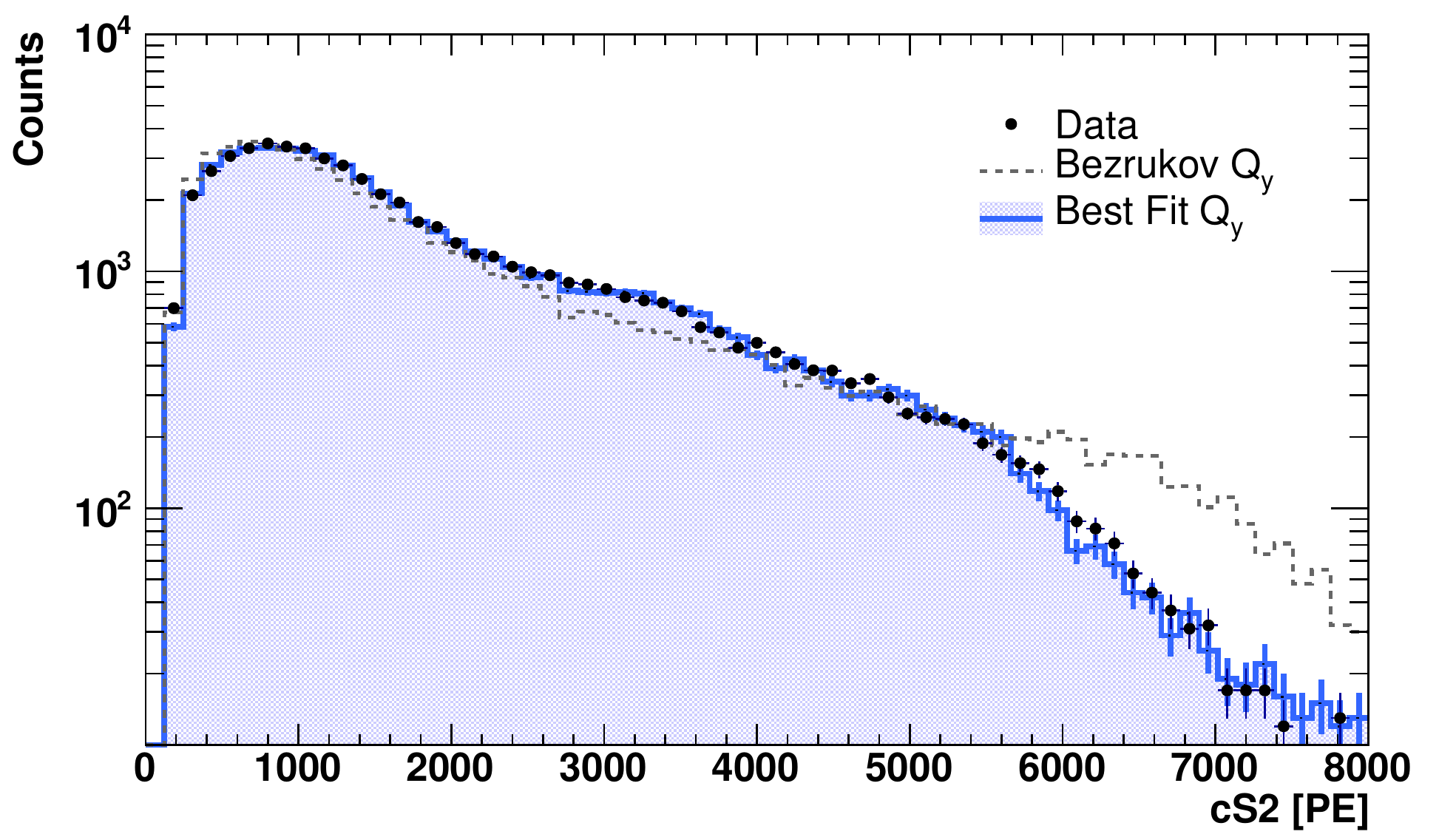}
\caption{\label{s2_matching}  
Comparison between the MC and data \cStwo\ spectra. The black data-points indicate the data and the blue spectrum is obtained as the result of the optimization of \qy. Good agreement between spectral shape and absolute rate across the whole signal range is achieved. 
For comparison, the gray dashed line indicates a generated \cStwo\ spectrum, assuming the same \qy\ as shown by the dashed line in Fig.~\ref{qy_bestfit} and described in Ref.~\cite{bezrukov2010}. }
\end{figure}

\begin{figure}[h]
\includegraphics[width=.5\textwidth]{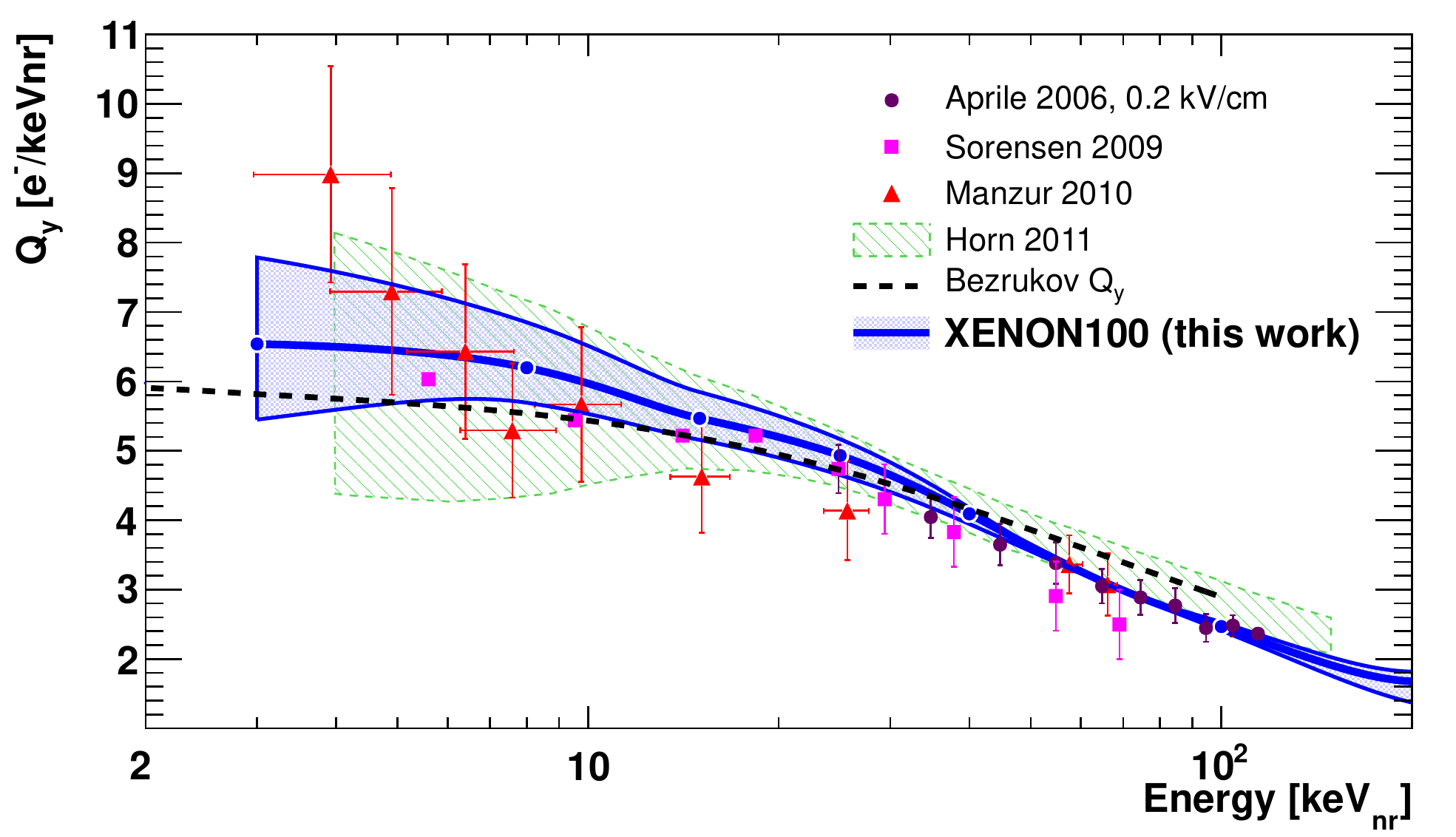}
\caption{\label{qy_bestfit} 
Result on \qy\ obtained from fitting the MC generated \cStwo\ spectrum to data. Pivot points of the spline interpolation are shown in light blue. The shaded area indicates the systematic uncertainty from varying input parameters of the simulation (find discussion in text). The interpolation between the pivot points at 0.5 and 3~\kevr\ does not yield a reliable result for \qy\ and is therefore not shown. The purple data points show the result of the first measurement of \qy\ in LXe at 0.2~kV~cm$^{-1}$~\cite{aprile2006}. Red data points show the result from direct measurements at a drift field of 1.0~kV~cm$^{-1}$~\cite{manzur2010}. The green hatched area and magenta data points are the combined first and second science run result from the ZEPLIN-III experiment~\cite{horn2011} and the result from the XENON10 experiment~\cite{sorensen2009}, respectively. Both results were extracted in a similar fashion to this work although the ZEPLIN-III parametrization was derived from data taken at a much higher field. The black dashed line represents a predicted \qy\ based on a specific phenomenological model as described in Ref.~\cite{bezrukov2010}. }
\end{figure}

The resulting pivot points and systematic errors together with the spline interpolation yield a best-fit \qy\ function. 
Fig.~\ref{s2_matching} shows the spectral matching corresponding to the central fit value of \qy\ (shown in Fig.~\ref{qy_bestfit}) along with data. The spectral matching is good from low to high values of \cStwo.
In Fig.~\ref{qy_bestfit} \qy\ is determined down to $\sim$3~\kevr\ with similar precision as achieved in the direct measurement~\cite{manzur2010} but, in
the lowest energy intervals, the central curve obtained here shows a trend which is flatter compared to Manzur et al. It is, however, still compatible within errors.
The uncertainty on \qy\ reduces with increasing recoil energy as the impact of variations in acceptance and \leff\ diminish. 
Best matching is achieved with a simulated neutron emission rate of 159~n/s which is in agreement with the independently measured neutron emission rate of $(160 \pm 4)$~n/s as described in Sec.~\ref{sec:modellingneutrons}.

\subsection{Scintillation Channel -- Determining \leff}

As with the extraction of the best-fit \qy\ in the previous sub-section, the \leff\ fit is parameterized by the same 8 independent spline points, which are allowed to vary to give the best agreement between data and simulated \cSone\ spectrum. The fit range is constrained to 2-200~PE as good agreement below 2~PE has not been achieved for a wide variety of \leff\ parameterizations. This mismatch is predominantly due to uncertainties in the calculated efficiency curve as given in Ref.~\cite{aprile2012} but it may, in part, also be due to uncertainties in the neutron physics provided by GEANT4 at the lowest recoil energies. A further cause of uncertainty is that the calibration of the detector response becomes difficult as signals approach the single- or two-photoelectron level where the PMT response due to single photoelectron size noise or electronics noise becomes more difficult to characterize.

For the simultaneous generation of \Stwo\ as a part of the complete signal simulation, the central curve of the previous fit result for \qy\ is used. Consequently, any result on the best-fit \leff\ has a systematic uncertainty of the size of the $\pm1\sigma$ bounds of the \leff\ representation in Ref.~\cite{aprile2011}, which was found to be the main contributor for the estimated error on \qy\ as presented in Fig.~\ref{qy_bestfit}. 

\begin{figure}[h]
\includegraphics[width=.5\textwidth]{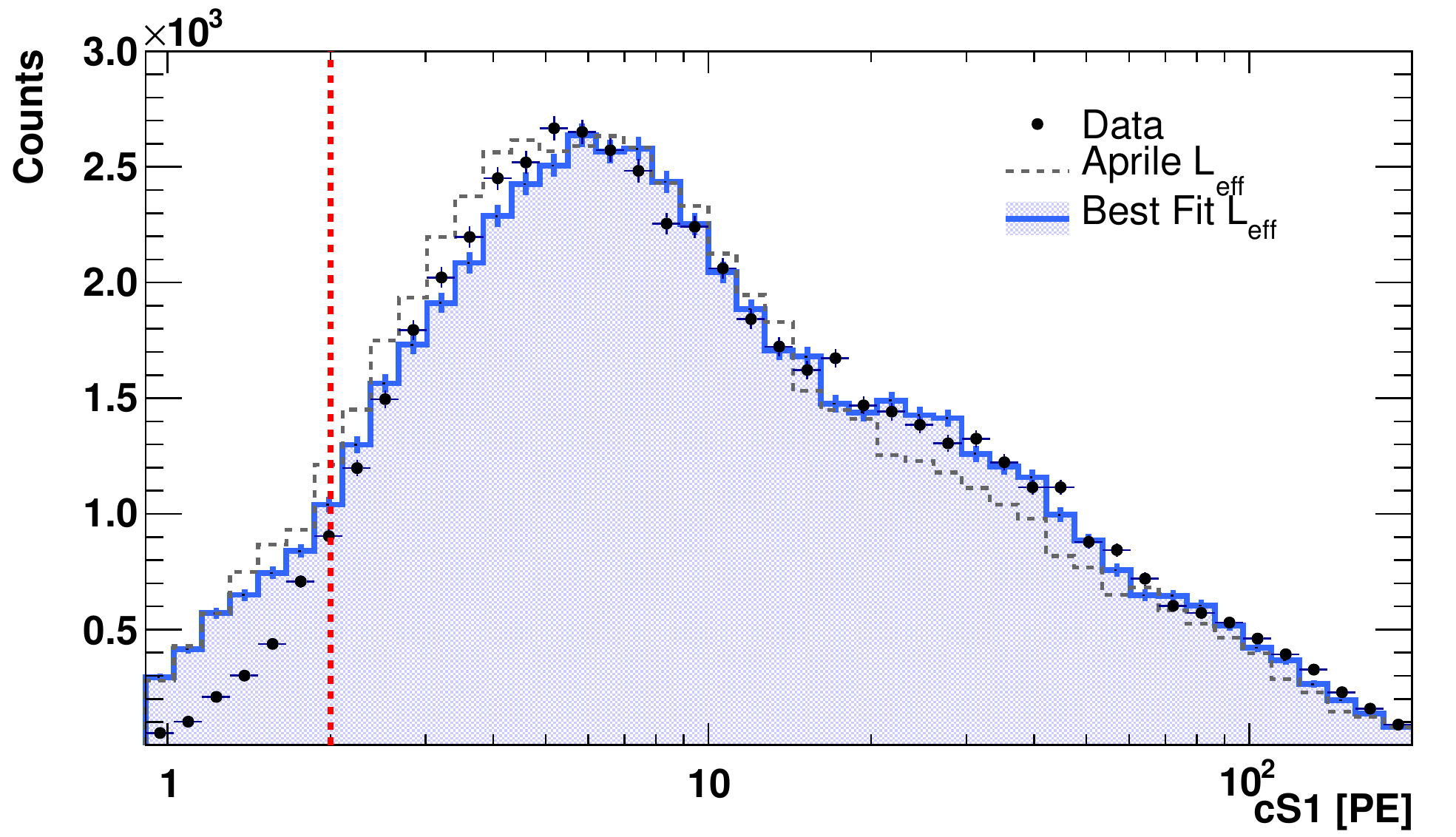}
\caption{\label{s1spec} 
Fit of the simulated \cSone\ spectrum to data (black points). The MC spectrum (blue) is obtained using the \leff\ after the optimization process. Below 2~PE (indicated by the vertical red line), a discrepancy between data and MC is observed (see text for discussion). The gray dashed line shows the spectral shape using the \leff\ detailed in Ref.~\cite{aprile2011} for comparison. Reasonable agreement between data and MC above 2~PE is already achieved with this \leff\ parameterization.}
\end{figure}

\begin{figure}[h]
\includegraphics[width=.5\textwidth]{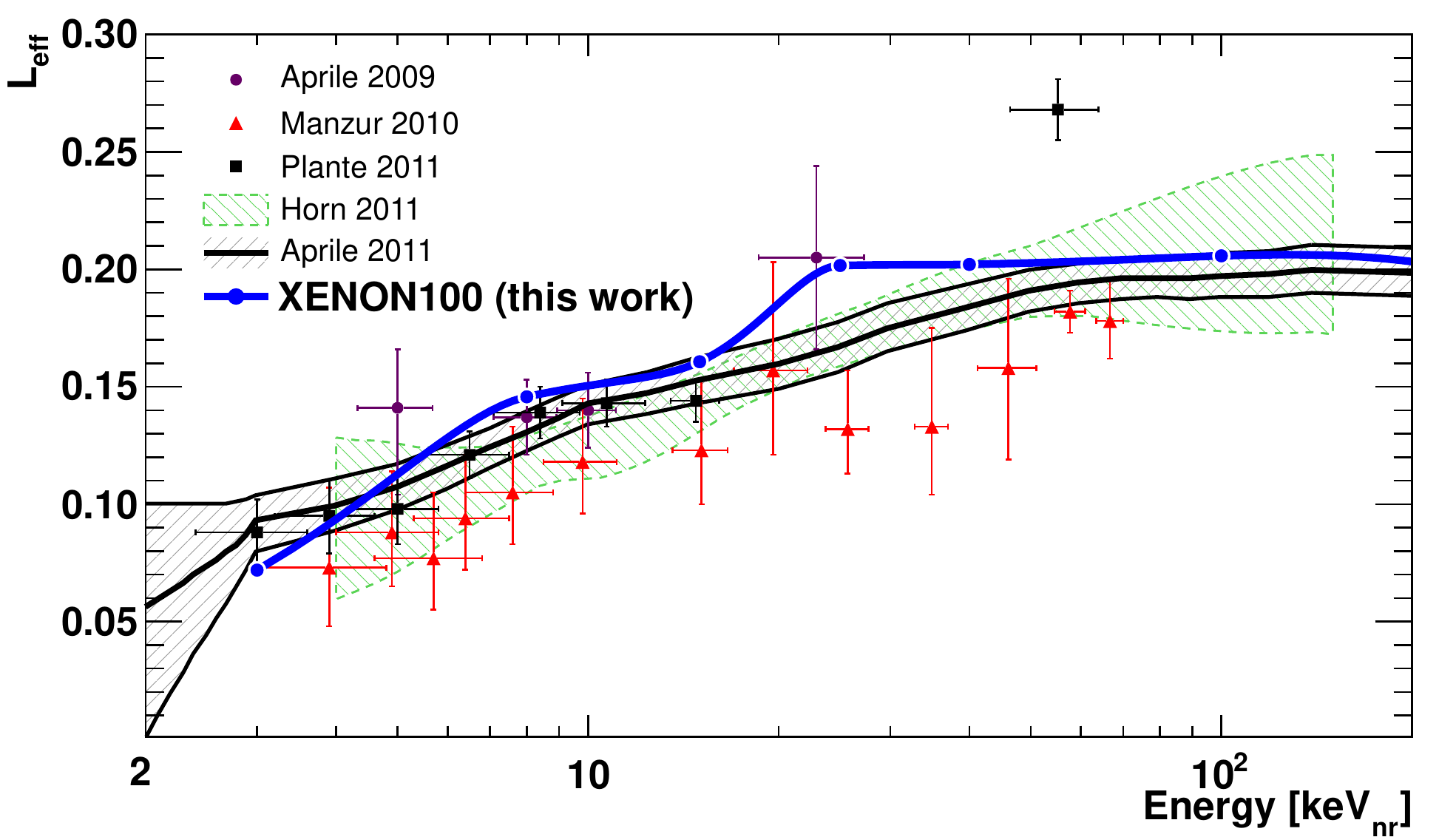}
\caption{\label{leff} 
\leff\ (blue line) obtained after the optimization of the absolute \cSone\ matching. As with Fig.~\ref{qy_bestfit}, pivot points of the spline interpolation are shown in light blue. As with the extraction of \qy, the parameterization of \leff\ is unreliable below 3~\kevr\ and is therefore not shown. For comparison, literature values of \leff\ including Aprile $et~al.$ $({\CIRCLE})$~\cite{aprile2009}, Manzur $et~al.$ $({\triangle})$~\cite{manzur2010}, Plante $et~al.$ $({\blacksquare})$~\cite{plante2011}, Horn $et~al.$ combined result (green-shaded)~\cite{horn2011} are shown along with this work (blue). Also shown is a global fit to all \leff\ data, used in Ref.~\cite{aprile2011} (black line and gray-shaded uncertainty). }
\end{figure}

Fig.~\ref{s1spec} presents the result of matching for the \leff\ (gray) as described in Ref.~\cite{aprile2011} and applied at the beginning of the fitting process while the best-fit \leff\ line is represented by the blue spectrum. The optimized \leff\ is shown in Fig.~\ref{leff} in comparison with the literature values.

It is shown that this \leff\ is in good agreement with that measured by Plante $et~al.$~\cite{plante2011} and in good agreement overall with other measurements below 15 \kevr. The deviation between the extracted \leff\ and the mean measurement results from the improvement of the spectral matching in the range of $20-60$ PE in Fig.~\ref{s1spec}.

The mean extracted \leff\ provides an important consistency check to recent direct measurements but the method does not improve on the accuracy on this quantity until more precise direct measurements of Qy as a function of energy and drift field become available.

\subsection{Two-Dimensional Distributions}

When satisfactory data/simulation agreement is achieved for the ionization and scintillation channels individually, the combined two-dimensional distributions provide a robust test of the consistency of the derived nuclear recoil energy scales.  
This is because even though individually the \cSone\ spectrum is only weakly sensitive to changes in \qy\ and the \cStwo\ scale only weakly sensitive to changes in \leff, a two-dimensional comparison is sensitive to both. 

\begin{figure}[h]
\includegraphics[width=.5\textwidth]{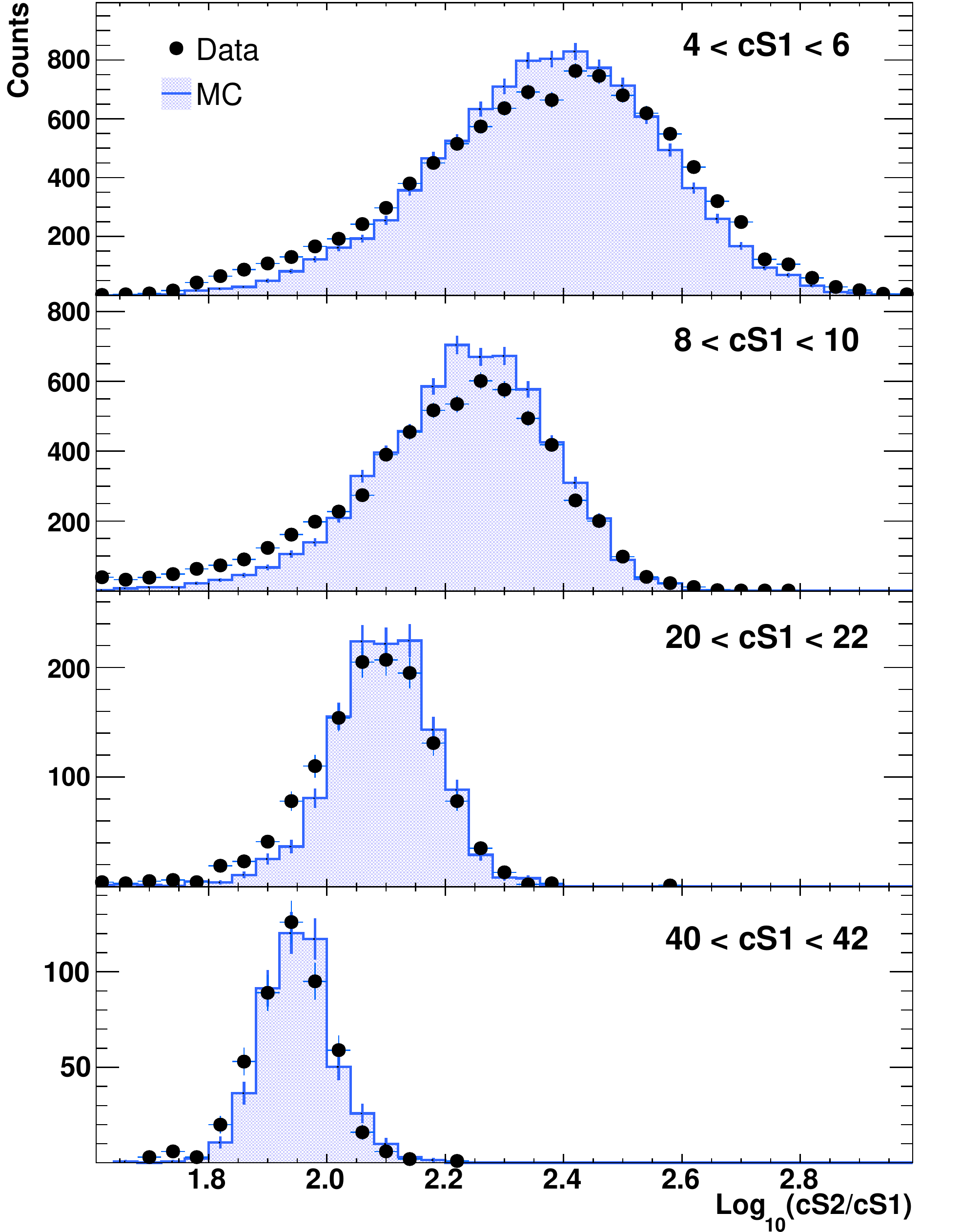}
\caption{\label{slices}
Projection of log$_{10}$(\cStwo/\cSone) in data (black) and MC (blue) for slices in \Sone. Good agreement is found for all slices. Mild deviations are observed at lower values of log$_{10}$(\cStwo/\cSone). A possible reason for these deviations is discussed in the text.}
\end{figure}

The log$_{10}$(\cStwo/\cSone) projections of the two-dimensional distributions for both data and MC are sliced into 2~PE bins and compared. Fig.~\ref{slices} shows several of these distributions.
Upon examination, it is clear that the matching between data and MC is reasonable but there are some variations, particularly at the lowest values of log$_{10}$(\cStwo/\cSone). 
This variation is most likely due to the presence of anomalous events containing two recoil vertices (\cSone\ and \cSone$^\prime$), one of which (\cSone$^\prime$) occurs in a region where the electric field configuration will not allow ionization electrons to be drifted to the liquid/gas interface. 
The recoil which occurs in such a region will have no associated ionization signal and, as such, is indistinguishable from a single scatter event (for electronic recoil backgrounds, this effect has been reported initially in Ref.~\cite{angle2008} and subsequently in Ref.~\cite{lebedenko2009}). 
These signals will have a relatively larger \cSone\ signal as compared to that expected given the size of the \cStwo\ signal (actually associated to only one of the two recoils) causing the apparent ratio of \cStwo/\cSone\ to fall.
The simulation includes events where the second recoil occurs between the cathode and the lower PMT array. 
The \cSone$^\prime$ signal is calculated using a LCE map which has been extrapolated below the cathode. 
Any remaining variations can be attributed to recoil events where the \cSone$^\prime$ recoil occurs in a region where the LCE is not precisely known (such as between PMTs in the lower array) and cannot be predicted.

\begin{figure}[h]
\includegraphics[width=.5\textwidth]{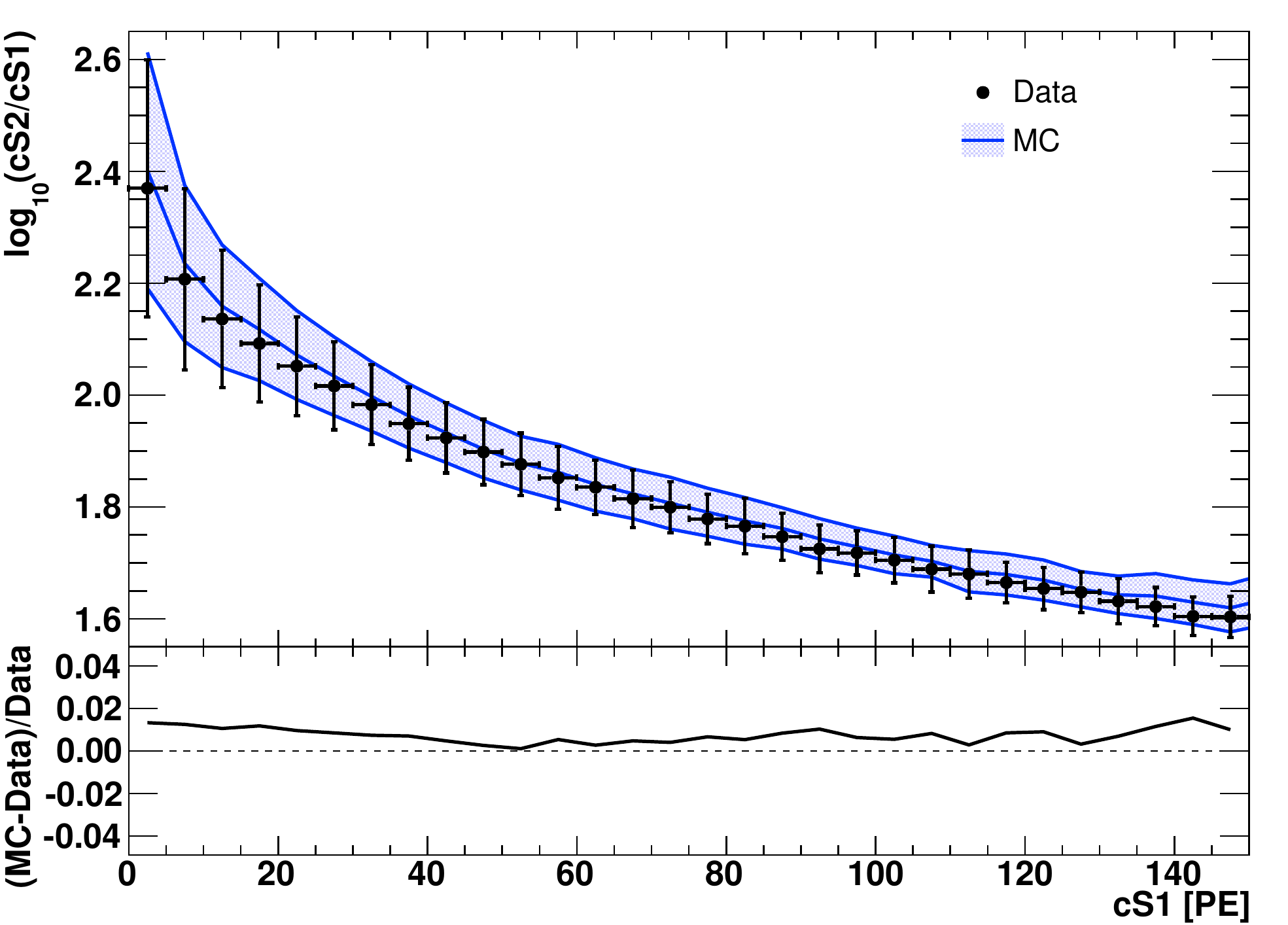}
\caption{\label{2dag}
Two-dimensional distributions of the means of the single scatter elastic nuclear recoil population from $^{241}$AmBe calibration data (black) and MC simulations (blue), where log(\cStwo/\cSone) is the discrimination parameter and the energy scale is defined by the primary scintillation channel (top). The error bars (black) and filled area (blue) represent the $\pm 1\sigma$ bands for data and MC respectively. The bottom panel shows the residual differences between data and MC which are all within 2\%.}
\end{figure}

Fig.~\ref{2dag} shows the parameterization of the elastic nuclear recoil band from the $^{241}$AmBe calibration in data and MC in the phase-space of log$_{10}$(\cStwo/\cSone) against \cSone. Points represent the medians of the distribution intervals in energy as defined by the scintillation channel. 
Vertical bars represent the $\pm1\sigma$ quantiles of each slice, encompassing the statistical fluctuations in signal generation and detector resolution.
Good agreement is achieved between calibration data and MC for both means and widths of each distribution.
Although the log$_{10}$(\cStwo/\cSone) distribution is systematically higher for the MC, it is still within 2\% of the central data values.
In this 2-dimensional distribution, the effect seen at low values of log$_{10}$(\cStwo/\cSone) in Fig.~\ref{slices} will cause events to move to log$_{10}$(\cStwo/(\cSone+\cSone$^\prime$)) whilst simultaneously moving to a higher apparent value of \cSone\ (namely \cSone+\cSone$^\prime$). The most pronounced changes will occur when \cSone$^\prime$ is of a similar or greater size than \cSone.
The level of agreement between data and MC in this comparison clearly shows that the effect of multiple scatter events that appear consistent with a single scatter event is minimal (as the maximum observed deviation is  $<2\%$). 
  
\begin{figure}[h]
\includegraphics[width=.5\textwidth]{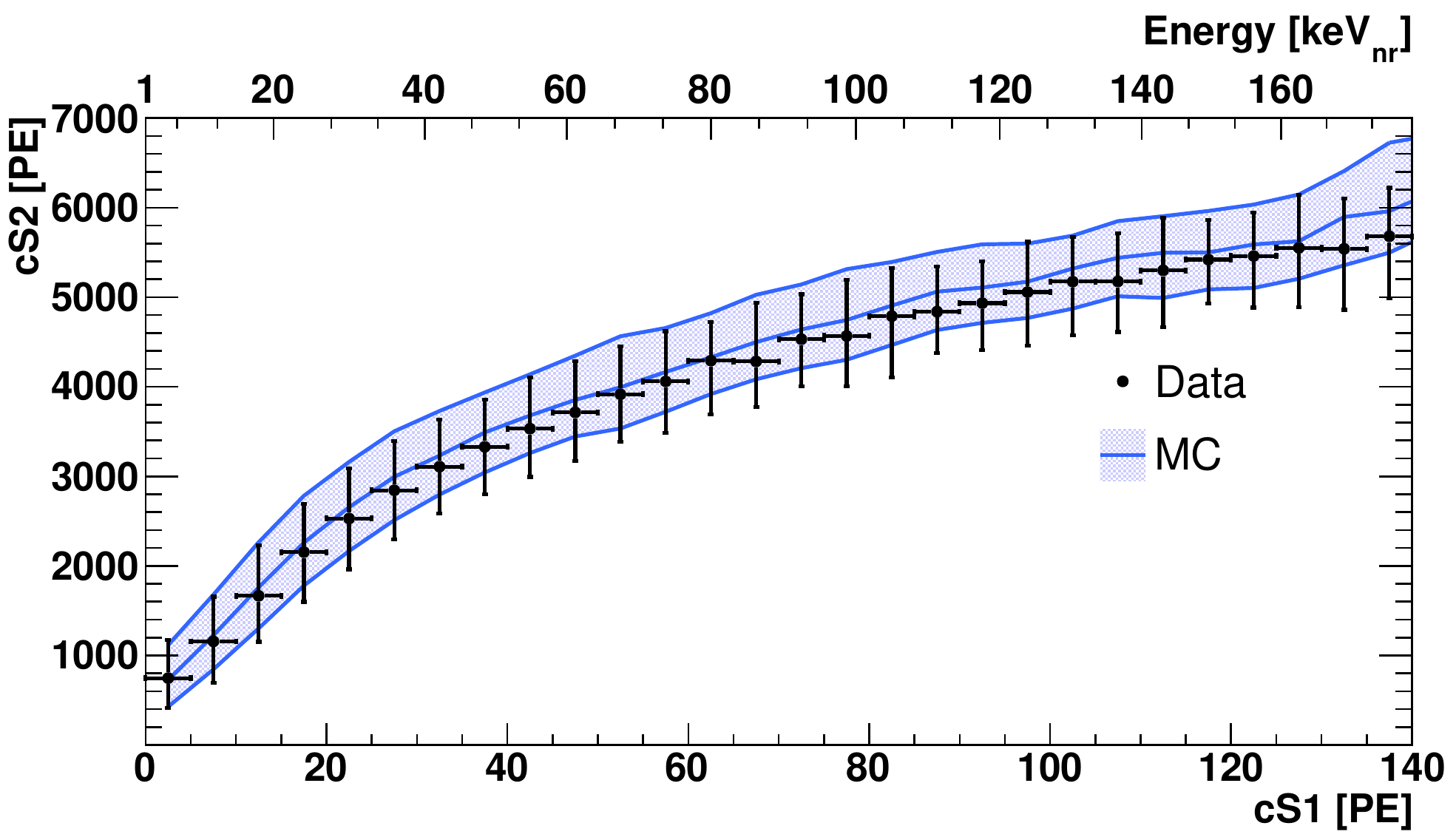}
\caption{\label{s1s2}
Comparison of \cStwo\ vs \cSone\ for data (black) and MC (blue). The solid lines represent the median of the \cStwo\ vs \cSone\ distribution while the vertical lines (black) and filled area (blue) represent the $\pm 1\sigma$ bands for data and MC respectively. As with the parameterization reported in~\cite{aprile2011}, the extracted \leff\ infers a energy scale that reaches 1~\kevr\ at 0~PE.}
\end{figure}

In addition to comparing log$_{10}$(\cStwo/\cSone)~vs~\cSone\ for data and MC, a comparison in \cStwo\ vs \cSone\ is shown in Fig.~\ref{s1s2}. A good level of agreement is demonstrated over the full range in \cSone\ and \cStwo. 
Disagreement between data and MC increases at higher energies as the recoil energy spectrum from $^{241}$AmBe falls exponentially with recoil energy.

The level of matching achieved in the 2-dimensional comparison confirms that it is possible to reliably predict nuclear recoil event signatures in the both signal channels simultaneously. 

\section{Simulated WIMP Distribution}
\label{sim_WIMP}

Simulating the response of \xehund~in both ionization and scintillation channels gives the ability to predict the distribution of WIMP recoil events in the discrimination parameter space.
An isothermal WIMP halo with a local density of $\rho_{\chi} = 0.3~\mathrm{GeV}$~cm$^{-3}$, a local circular velocity of $\nu_{0} = 220~\mathrm{km}~\mathrm{s}^{-1}$, and a Galactic escape velocity of $544~\mathrm{km~s}^{-1}$~\cite{smith2007} is assumed.
For the calculation of expected rates, \leff\ from Ref.~\cite{aprile2011} and \qy\ as determined in this work are used.

\begin{figure}[h]
\includegraphics[width=.5\textwidth]{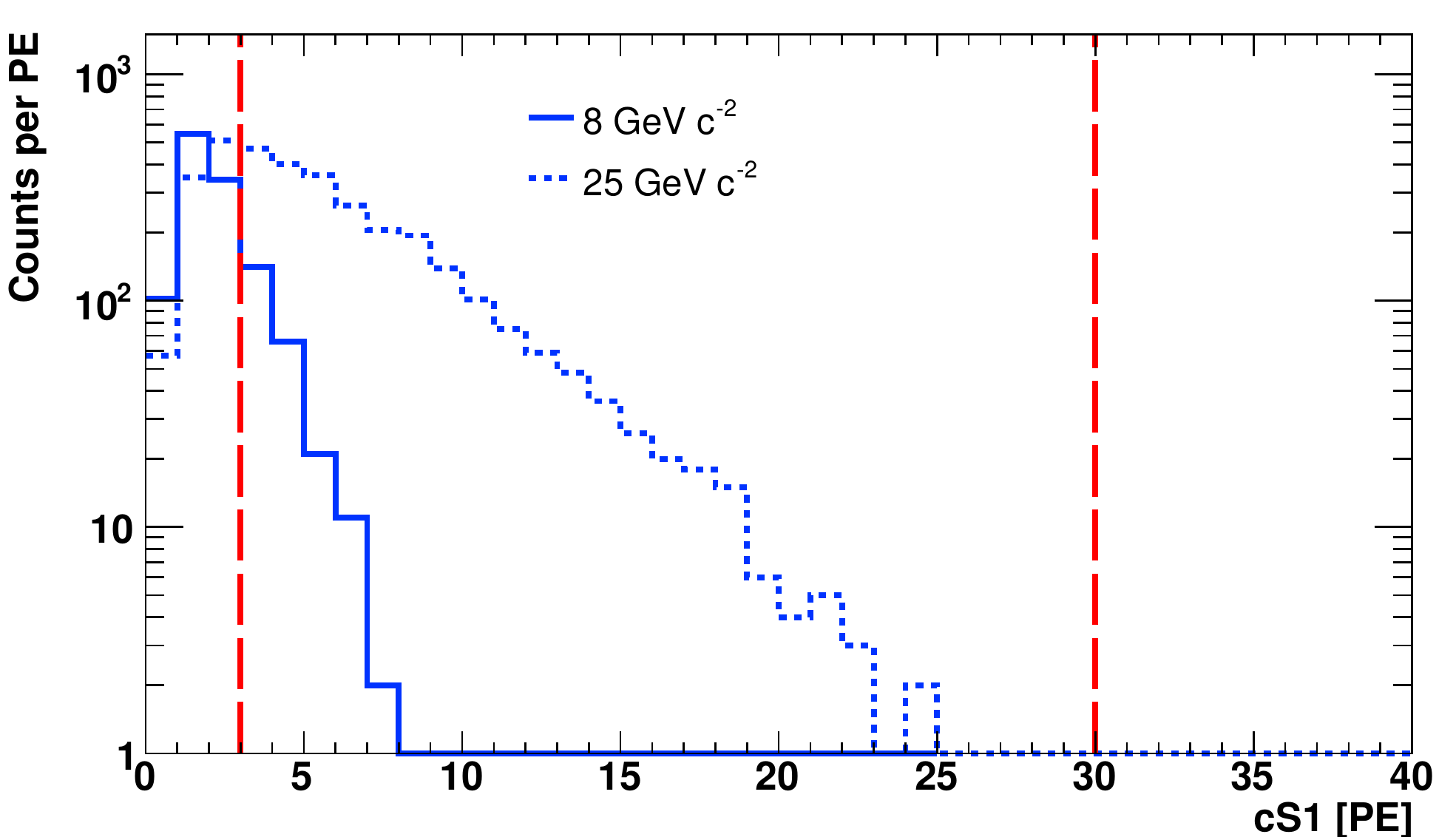}
\caption{\label{1D_WIMP}
Recoil spectra for (solid blue) an $8~\mathrm{GeV}~c^{-2}$ WIMP and (dashed blue) a $25~\mathrm{GeV}~c^{-2}$ WIMP with spin-independent  WIMP-nucleon cross-sections of $3\times10^{-41}~\mathrm{cm}^{2}$ and of $1.6\times10^{-42}~\mathrm{cm}^{2}$, respectively, as they would be observed in \xehund\ taking into account energy resolution and detection efficiencies. Boundaries of 3 and 30 PE in \cSone\ are marked by the dashed red lines. In both cases, standard WIMP parameters are assumed (see Sec.~\ref{sim_WIMP}) and an exposure equivalent to that of the 225 live-days \xehund~WIMP search run with a 34~kg fiducial mass is used.}
\end{figure}

Sample recoil spectra for both an $8~\mathrm{GeV}~c^{-2}$ WIMP mass with a spin-independent WIMP-nucleon cross-section of $3\times10^{-41}~\mathrm{cm}^{2}$ and for a $25~\mathrm{GeV}~c^{-2}$ WIMP mass with a spin-independent WIMP-nucleon cross-section of $1.6\times10^{-42}~\mathrm{cm}^{2}$ are shown in Fig.~\ref{1D_WIMP}.
WIMP masses are chosen such that the signal response to both light and more massive particles is investigated. The choice of the interaction cross-sections is motivated by the intersection of these masses with the lower boundary (central region) of recent dark matter signal claims from the CoGeNT~\cite{aalseth2011} and CRESST-II~\cite{angloher2011} experiments.    
The distributions are generated for a dark matter exposure equivalent to that of the 225 live-days \xehund~dark matter search. 
From the given recoil spectra, simulated \cSone\ and \cStwo\ signals are generated whilst maintaining the simulated cuts described in Sec.~\ref{cuts}.
The efficiency for all data quality cuts is also applied.

\begin{figure}[h]
\includegraphics[width=.5\textwidth]{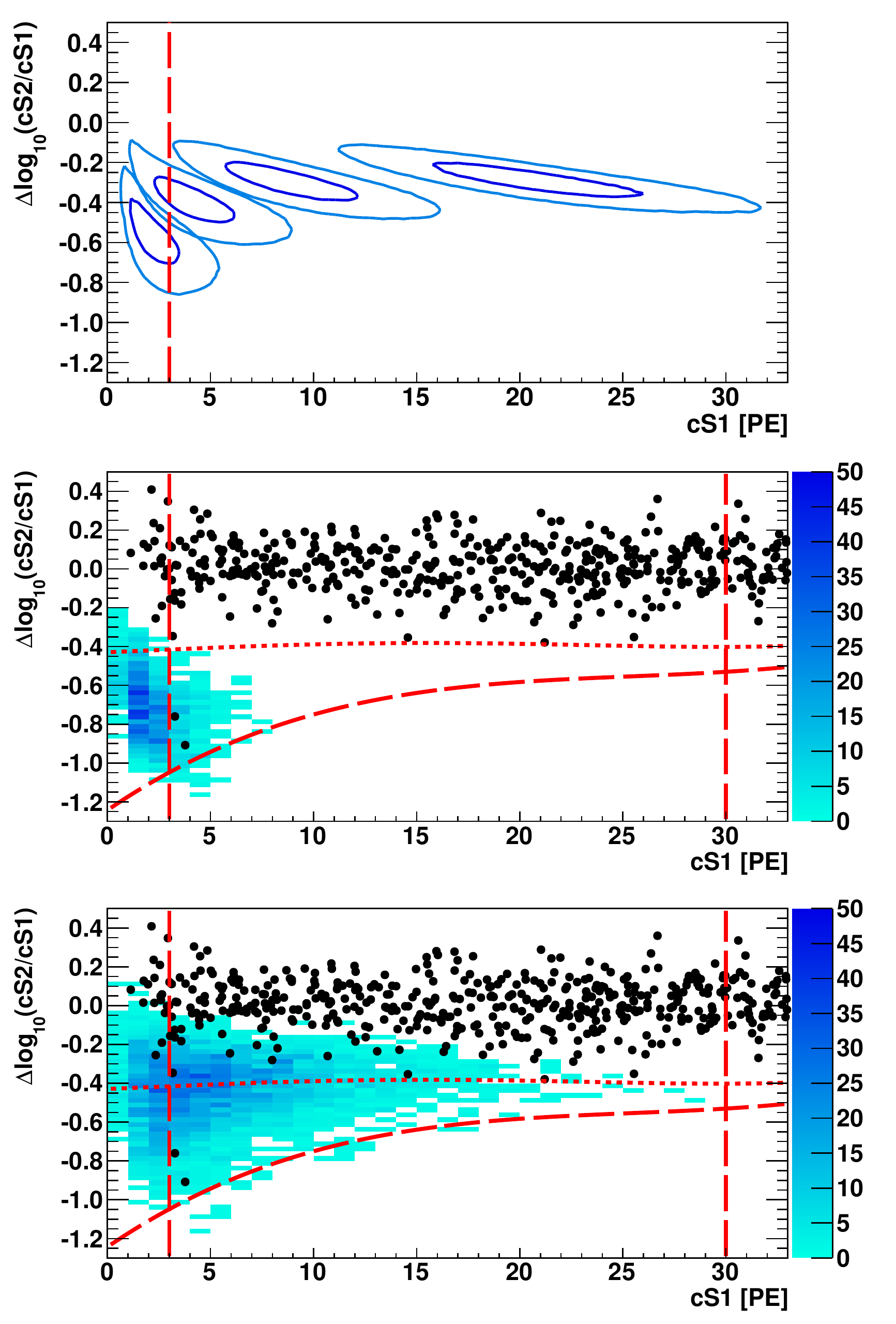}
\caption{\label{2D_WIMP}
Two-dimensional distributions of expected \cSone\ and \cStwo\ signals for (top) mono-energetic nuclear recoils of 4, 8, 16 and 32 \kevr\ (represented using 1$\sigma$ and 2$\sigma$ contours), (middle) for an $8~\mathrm{GeV}~c^{-2}$ WIMP and for (bottom) a $25~\mathrm{GeV}~c^{-2}$ WIMP with spin-independent  WIMP-nucleon cross-sections of $3\times10^{-41}~\mathrm{cm}^{2}$ and of $1.6\times10^{-42}~\mathrm{cm}^{2}$, respectively. The same assumptions used to generate the recoil spectra in Fig~\ref{1D_WIMP} were used. The vertical red lines represent the 3 PE analysis threshold and the upper 30 PE boundary (lower 2 panels).
in the lower two panels, the horizontal (long-dash) red curve represents the mean ($\mu$) $- 3\sigma$  for the elastic nuclear recoil distribution and the horizontal (short-dash) red curve represents the 99.75\% electron recoil rejection line as discussed in Ref.~\cite{aprile2012}.}
\end{figure}

Fig.~\ref{2D_WIMP} shows simulated distributions in discrimination parameter space for (top panel) several values of monoenergetic nuclear recoils, showing the expected spread caused by fluctuation in cS2 and cS1 along with two WIMP masses (middle and bottom panels) superimposed on the \xehund\ 225 live-days dark matter search data for comparison~\cite{aprile2012}.
The distributions are presented in a parameter space which is flattened by subtracting the mean of the electronic recoil distribution.
As expected, the recoil spectrum of a $25~\mathrm{GeV}~c^{-2}$ WIMP gives rise to a larger number of events extending to higher recoil energies than the $8~\mathrm{GeV}~c^{-2}$ WIMP and its bulk distribution resembles very well the shape of broad-band $^{241}$AmBe neutron calibration.
For the lower mass WIMP, the majority of events are expected below the analysis threshold in a region where \xehund\ is still sensitive, but a significant number of recoil events fall above the imposed analysis threshold of 3~PE.  However, the center of the signal distribution is clearly shifted towards a lower log$_{10}$(\cStwo/\cSone) with respect to the average neutron band position. This can be explained by the steeply falling recoil energy spectrum of light mass WIMPs combined with asymmetric upward fluctuations in the Poisson dominated regime of small S1. 
For both cases the expected number of WIMPs is calculated for an exposure equivalent to the 225 live-days \xehund\ WIMP search run in a region given by a S1 range of 3-30 PE S1 and below the 99.75\% electron recoil line as defined in Ref.~\cite{aprile2012}. The results are $223^{+303}_{-85}$ (sys.) and $1409 ^{+53}_{-4}$ (sys.) events for the $8~\mathrm{GeV}~c^{-2}$ and for the $25~\mathrm{GeV}~c^{-2}$ WIMP, respectively.
In both cases, statistical errors are subdominant as the distribution is created using large numbers of statistics and is scaled to the calculated exposure. The systematic error is defined by simultaneously using the upper and lower bounds of \leff\ and \qy.
Rates could similarly be calculated for the \leff\ extracted in this publication. The shape of this Leff leads to predicted rates consistent with those calculated for the direct Leff measurements within errors.

The excess of predicted WIMP recoil rates above only 2 event candidates observed in the 225 live-days \xehund\ dark matter search~\cite{aprile2012} is consistent with the reported exclusion limit, supporting the tension between these results and signal claims by other experiments \cite{aalseth2011,dama,angloher2011}.

\section{Conclusions}

The neutron calibration of the \xehund\ dark matter detector with a $^{241}$AmBe source has been modeled with a MC simulation that includes the signal generation in both the \Sone\ and \Stwo\ channels. 
Agreement in the ionization channel is achieved through the adoption of a \qy (derived using $^{241}$AmBe data and a fixed \leff) that is largely consistent with previous direct and indirect measurements and phenomenological estimations but shows no indication of a low-energy increase as reported by the direct measurement of Ref.~\cite{manzur2010}. Additionally, an optimized \leff\ is determined using a similar method and is used to match data and MC signal distributions in the scintillation channel. 
The ionization and scintillation channels are combined in two-dimensional spaces, achieving agreement between MC and data, constraining the uncertainty in the nuclear recoil energy scales, and reproducing both means and widths of energy distributions. It provides a strong validation of the understanding of the discrimination parameter space in which previous \xehund\ dark matter searches were analysed and reported. 
A simulated neutron emission rate of 159~n/s is required to achieve spectral matching. This is in agreement with the measured emission rate of $(160 \pm 4)$~n/s and confirms the robustness of the S1 signal acceptance used in the \xehund\ WIMP searches~\cite{aprile2011,aprile2012,aprile2012_anl,aprile2012_SI}.

We gratefully acknowledge support from NSF, DOE, SNF, UZH, FCT, INFN, R\'{e}gion des Pays de la Loire, STCSM, NSFC, DFG, Stichting voor Fundamenteel Onderzoek der Materie (FOM), the Max Planck Society, the Weizmann Institute of Science and the EMG research center. We are grateful to LNGS for hosting and supporting \xehund. 

\bibliographystyle{apsrev}
\bibliography{XE100AmBePaper}

\end{document}